\documentclass[12pt]{article}
\usepackage[utf8x]{inputenc} 
\usepackage{amssymb}
\usepackage{mathrsfs}
\usepackage{bm}
\usepackage{graphicx}
\usepackage{subfigure}
\usepackage{hyperref}
\usepackage{cite}
\begin{document}
\begin{titlepage}
\begin{flushright}
\par\noindent BiBoS 506/92\\
\end{flushright}
\begin{center}
\begin{large}
{\bf Dynamics of Contact Epidemic Models for Finite Populations}
\end{large}
\vskip1truecm
Ph. Blanchard$^{(a)}$ and Stam Nicolis$^{(b)}$

${ }^{(a)}${\sl BiBoS Forschungszentrum, Universität Bielefeld\\ Universitätsstraße 25, 4800 Bielefeld 1, Germany}

\vskip0.5truecm

${ }^{(b)}${\sl Höchstleistungsrechenzentrum der KFA Jülich\\ Postfach 1913, 5170 Jülich 1, Gemany}
\end{center}

\vskip1truecm

\begin{abstract}
We study contact epidemic models for the spread of infective diseases in finite populations. The size dependence enters in the infection rate. The dynamics of such models 
is then analyzed within the deterministic approximation, as well as in terms of a stochastic formulation. At the level of the deterministic equations we deduce relations between the parameters of the model for the disease to become endemic. Within the stochastic formulation it is possible to write recursion relations for the probability distribution, that lead to exact expressions for some of its moments and check the range of validity of the stochastic threshold theorems. 
\end{abstract}
\end{titlepage}
\newpage
\section{Introduction}\label{intro}
Contact epidemic models arise in a variety of situations in biology, chemistry and physics: e.g. in genetics and ecology~\cite{bailey1975mathematical} (from where the term originated), in chemical kinetics~\cite{nicolis1972fluctuations,MalekMansour1973} and non-equilibrium statistical mechanics~\cite{nicolis1977self,van1984stochastic,bartelt1991kinetics}. Recently they have been studied within the framework of directed percolation~\cite{privman1991unusual,privman1991infinite,aizenman1991strict}. 

The present study addresses the problem of propagation of an epidemic in a finite population. The issue here is under what circumstances a disease may become endemic. Assuming that the infection rate depends on the population size in a certain way, we attempt a study of the possible consequences. The deterministic approximation is the first step, where we apply elementary ideas from dynamical systems theory and derive bounds on the population size and on other parameters  for the disease to become endemic
and for the constraints to be meaningful. Within the stochastic formulation we present a computer--aided solution for the general stochastic epidemic as well as for the opposite case, where susceptibles may enter the population, but infectives do not leave. Our conclusions and further discussion are the subject of the last section. 

\section{The Deterministic Approximation}\label{detapprox} 
Deterministic models of infective diseases are based upon the law of mass--action of chemical kinetics: the rate of change of the number of susceptibles (or infectives) is proportional to the product of the current number of susceptible and infected individuals. 

In fact matters are considerably more complicated. A complete model should take into account the birth rate of susceptible individuals, the death rate of susceptibles, infectives and immune and the fact that infected individuals may become immune or get well (reenter the class of susceptibles). All this may be written, within the deterministic approximation, in the form of rate equations for the evolution in time of the number of susceptible $X$, infected $Y$ and immune $Z$ individuals as follows
\begin{equation}
\label{SIRmodel}
\begin{array}{lcc}
\displaystyle
dX/dt & = & \displaystyle \zeta(1+b(N))^{-1}N-\mu(1+m(N))X-\beta(N)XY+(1-p)\gamma Y\\
\displaystyle 
dY/dt & = & \displaystyle \beta(N)XY-(\gamma+\mu(1+m(N)))Y\\
\displaystyle
dZ/dt & = & \displaystyle p\gamma Y -\mu(1+m(N))Z
\end{array}
\end{equation}
$\zeta$ is the rate, with which susceptibles enter the population and contains a term that depends on the total population size $N$ (the term $1/(1+b(N))$); $\mu$ is the rate, with which susceptibles, infectives and immune leave the population (a death rate); it also receives a corrective factor, $(1+m(N))$, that takes into account the finite size of the population. The infection rate is $\beta(N)$. Finally, infected individuals may become immune at a rate $p\gamma$ and susceptible at  a rate $(1-p)\gamma$. Since an individual must  belong to one of these classes, the total population, $N=X+Y+Z$ is constant; this leads to a constraint on the parameters of the model
\begin{equation}
\label{zetaconstraint}
\zeta = \mu(1+m(N))(1+b(N))
\end{equation}
The validity of this model rests on the choice of the functions $b(N), \beta(N)$ and $m(N)$ and on the values of the other parameters. A complete study is very difficult so, in what follows, we shall focus on certain aspects, in an effort to clarify the assumptions involved. 

First of all, we shall consider the dynamics in the subspace of the susceptibles $X$ and the infectives $Y$ only. We shall try to compare the resulting model with a Lotka--Volterra system, whose dynamics has been extensively studied within the same context, in order to see the similarities and differences in behavior between them. 

\subsection{The Lotka--Volterra Model}\label{LVmodel}
Taking into account a renewal of the susceptible population and a removal rate of the infected individuals, one may write the equations for the {\em densities} of the susceptibles $x=\#\mathrm{susceptibles}/N$ and the infected $y=\#\mathrm{infected}/N$ (where $N$ is the initial size of the total population) as follows
\begin{equation}
\label{LVeqs}
\begin{array}{lcc}
\displaystyle
dx/dt & = &\displaystyle -\beta xy +\mu x(1-\nu x)\\
\displaystyle
dy/dt & = & \displaystyle \beta xy -\gamma y
\end{array}
\end{equation}
where $\beta$ is the {\em infection rate}, $\mu$ the {\em renewal rate} of the susceptible individuals, $\gamma$ the {\em removal rate} of the infected individuals and $\nu$ takes overpopulation into account (the second term of the first equation's rhs is nothing but the usual logistic term). 

These equations do not contain the population size $N$ at all; they may be considered valid for very large populations, where the densities may coincide with their average values. However an infection rate {\em does} depend on the population size, albeit weakly; if $\beta_\infty$ is the infection rate within an infinite population, it will be surely less for a finite one. How less may be expressed through the following empirical relation~\cite{schenzle1987critical}
\begin{equation}
\label{finitebetaN}
\beta(N)=\beta_\infty\frac{N}{\kappa + N}
\end{equation}
where $\kappa$ takes into account saturation within a finite population. It is clear that one may eliminate $N$ by rescaling $\kappa$
\begin{equation}
\label{rescalekappa}
\lambda = \kappa/N; \hskip0.5truecm \beta = \beta_\infty\frac{1}{1+\lambda}
\end{equation}
These manipulations are quite formal; one of the aims of the present study will be to try and impose restrictions on these parameters. 

To this end it is useful to realize that the eqs.~(\ref{LVeqs}) define a motion in the two--dimensional phase space $(x,y)$, given an initial composition $(x_0, y_0)$. Of great interest are asymptotic properties: how does the composition of the population behave for times long compared with the characteristic times of the system $(1/\mu, 1/\gamma)$? Does the disease become {\em endemic}--is $y(t\to\infty)\neq 0$? Or does it die out? Is the population as a whole wiped out by the disease? There exists a well--established body of theory that deals with these issues, the qualitative theory of ordinary differential equations~\cite{arnold1980equations}; it is quite straightforward to apply it here.

The asymptotic properties are encoded in the stationary configurations--those for which the rhs of~(\ref{LVeqs}) is zero. A simple computation gives three possible configurations
\begin{equation}
\label{stationaryconfigs}
\begin{array}{lcl}
\displaystyle x^\ast = 0, & & \displaystyle y^\ast = 0\\
\displaystyle x^\ast = 1/\nu, & & \displaystyle y^\ast = 0\\
\displaystyle x^\ast = \gamma/\beta, & & \displaystyle y^\ast = (\mu/\beta)\left(1-\nu\frac{\gamma}{\beta}\right)
\end{array}
\end{equation}
The first is the ``trivial"" state, where the population is wiped out. The second describes an ``immune'' state (a fraction of the susceptibles survives, but the infectives ar wiped out and the epidemic stops); the third describes, in general, an endemic state (since $y^\ast=0$ only for $\nu\gamma=\beta$). To determine in which of the three states the system will end up a linear stability analysis is carried out. 

\begin{itemize}
\item {\em Trivial State}. The eigenvalues of the stability matrix are $(\xi_1,\xi_2)=(\mu,-\gamma)$, indicating that this state is a {\em saddle point} and, consequently, unstable, independently of the parameter $\lambda =\kappa/N$, i.e. the population size. 
\item {\em Immune State}. The two eigenvalues are 
\begin{equation}
\label{immuneeigenval}
\begin{array}{lcl}
\displaystyle \xi_1 & = & \displaystyle \frac{1}{2}\left(\frac{\beta}{\nu}-(\mu+\gamma)\right)+\frac{1}{2}\sqrt{\left(\frac{\beta}{\nu}-(\mu+\gamma)\right)^2-4\mu\left(\gamma-\frac{\beta}{\nu}\right)}\\
\displaystyle \xi_2 & = & \displaystyle \frac{1}{2}\left(\frac{\beta}{\nu}-(\mu+\gamma)\right)-\frac{1}{2}\sqrt{\left(\frac{\beta}{\nu}-(\mu+\gamma)\right)^2-4\mu\left(\gamma-\frac{\beta}{\nu}\right)}\\
\end{array}
\end{equation}
Stability requires that both eigenvalues be negative, or (if they are complex) have negative real parts (it may be shown that this last case cannot occur). This leads to constraints on the population size:
\begin{equation}
\label{popsize1}
N\leq N_\mathrm{cr}=\kappa\frac{\nu(\mu+\gamma)}{\beta_\infty-\nu(\mu+\gamma)}
\end{equation}
and 
\begin{equation}
\label{popsize2}
N\leq N'_\mathrm{cr}=\kappa\frac{\nu\gamma}{\beta_\infty-\nu\gamma}
\end{equation}
It is interesting to note that these are {\em upper} bounds on the population size. It is, therefore, necessary to see what happens to the stability of the {\em endemic} state for such sizes.
\item {\em Endemic State}. The eigenvalues of the stability matrix are 
\begin{equation}
\label{enedemiceigenvalues}
\xi_{1,2}=-\frac{\mu\nu\gamma}{2\beta_\infty}\pm
\frac{1}{2}\sqrt{\left(\frac{\mu\nu\gamma}{\beta_\infty} \right)^2-4\gamma\mu\left(1-\frac{\nu\gamma}{\beta_\infty}\right)}
\end{equation}
in the complex case, therefore, the endemic state is stable; for real roots it is stable if the population size is {\em greater} than a critical value
\begin{equation}
\label{Ncritendemic}
N\geq N'_\mathrm{cr}=\kappa\frac{\nu\gamma}{\beta_\infty-\nu\gamma}
\end{equation}
which is one of the two critical sizes, that control the stability of the immune state!
\end{itemize}
The biological relevance of these configurations would require, however, that the values of $x^\ast,y^\ast$ lie between 0 and 1 (but see below).  This leads to further constraints on the population size (beyond the obvious one on the parameter $\nu$, viz. $\nu\geq 1$)
\begin{equation}
\label{Ncritdoupleprime}
N\geq N''_\mathrm{cr}=\kappa\frac{\gamma}{\beta_\infty-\gamma}
\end{equation}
from the relation $\gamma\leq\beta$; the relation $y^\ast\geq 0$ leads to $N\geq N'_\mathrm{cr}$. Finally, $y^\ast\leq 1$ is quadratic in $\lambda$. If $\mu < 4\nu\gamma$, the constraint is always satisfied; in the opposite case, 
\begin{equation}
\label{mugeq4nugamma}
\kappa\frac{\beta_2}{\beta_\infty-\beta_2}\geq N\,\mathrm{or}\,N\geq \kappa\frac{\beta_1}{\beta_\infty-\beta_1}
\end{equation}
where $\beta_{1,2}$ are given by 
\begin{equation}
\label{beta12}
\beta_{1,2}=\frac{\mu}{2}\pm\frac{\mu}{2}\sqrt{1-4\frac{\nu\gamma}{\mu}}
\end{equation}
There is a final constraint: $x^\ast + y^\ast$ must be less or equal to 1!! This translates into
\begin{equation}
\label{densitylt1}
N\geq N^\ast_\mathrm{cr}=\kappa\frac{\mu + \gamma}{\beta_\infty-\mu-\gamma}
\end{equation}
which is effective only if the infection rate is greater than the sum of the removal and the renewal rates. It should be noted, however, that, for the system under consideration, the population should fluctuate due to the influx of susceptibles and removal of infectives; only the initial condition should satisfy the constraint~(\ref{densitylt1}). 

It would be instructive to substitute `reasonable' values for the different parameters and see whether the constraints just derived are truly relevant for `real' systems. Even before doing so, however, we see that, if overpopulation is neglected, the immune state is irrelevant, while the endemic state is (in this case) always relevant and stable (to small fluctuations). Regarding {\em large} fluctuations, it is known that eqs.~(\ref{LVeqs}), for $\nu=0$ admit a conserved quantity~\cite{nicolis1972fluctuations,MalekMansour1973,nicolis1977self,frame1974explicit}
\begin{equation}
\label{conservedLV}
C(t)=\beta(x+y)-(\gamma\log x + \mu\log y)=C(0)
\end{equation}
If this curve is closed, the motion in the $(x,y)$ plane will be periodic, i.e. the endemic state will be stable to large fluctuations. It is possible to express the $x(t), y(t)$ and the period of the motion as power series~\cite{frame1974explicit}; the condition that the power series for the period converge is 
\begin{equation}
\label{periodconvseries}
\mathrm{max}\left\{\frac{\gamma C(0)}{4\pi},\frac{\mu C(0)}{4\pi}\right\} < 1
\end{equation}
which leads to a fairly complicated constraint involving the parameters of the model and the distance of the initial configuration to the endemic one. Unfortunately, this constraint is not at all tight. Of greater practical interest is the expression for the period, for which a very good approximation is~\cite{frame1974explicit}
\begin{equation}
\label{periodapprox}
T\approx\frac{4\pi}{\sqrt{\mu\gamma}}I_0\left(\sqrt{C(0)(\mu+\gamma)}/6\right)
\end{equation}
where $I_0(x)$ is the modified Bessel function. Since $I_0(x)$ is finite at $x=0$, the period diverges if either the removal rate of the infectives or the renewal rate of the susceptibles becomes very small. Hence, periodic solutions may be seen for values O(1) of these parameters. There is also the problem that, for too small values of $\mu$ or $\gamma$, the values of $x(t)$ or $y(t)$ escape from the allowed interval $[0,1]$. 

\begin{figure}[thp]
\begin{center}
\includegraphics[scale=0.6]{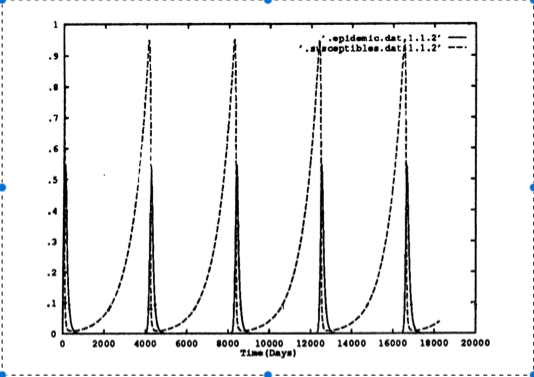}
\end{center}
\caption[]{\sl  Time evolution of susceptibles $x(t)$ (dashed line) and infectives $y(t)$ (solid line) for $1/\mu=75$ days, $1/\gamma=10$ days, $\beta_\infty=1$, $\kappa=1000$ and a population of $N=1000$.}
\label{Fig1}
\end{figure}

\begin{figure}[thp]
\begin{center}
\includegraphics[scale=0.6]{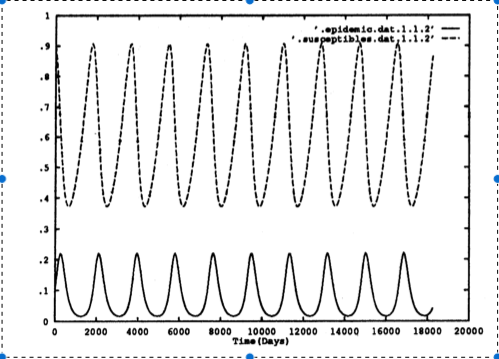}
\end{center}
\caption[]{\sl  Time evolution of susceptibles $x(t)$ (dashed line) and infectives $y(t)$ (solid line) for $1/\mu=75$ days, $1/\gamma=10$ days, $\beta_\infty=1$, $\kappa=1000$ and a population of $N=200$.}
\label{Fig2}
\end{figure}

In fig.~\ref{Fig1} we display numerical solutions for some typical parameter values. We estimate the period as $T\approx 4140$ days (vs. $T\approx 4380$ from eq.~(\ref{periodapprox})). Let us now comment on how effective the constraints on the population size are: in the case of fig.~\ref{Fig1} one finds that $N\geq N''_\mathrm{cr}=112$ and $N\geq N^\ast_\mathrm{cr}=128$, which indicates that the $\kappa-$factor plays a very important role for finite populations. Indeed, were $\kappa=O(1)$, then the constraints on the population size become completely ineffective. 
An interesting point for practical considerations is whether the epidemic will manifest itself as a series of large, infrequent, outbreaks, or small, frequent, ones and what will be the typical number of infected individuals between two outbreaks. The parameter values for fig.~\ref{Fig1} illustrate the first case; in fig.~\ref{Fig2} we display an example of the second, which indicates that populations, whose size is comparable to the saturation parameter $\kappa$ support large, infrequent, outbreaks, with quiescent periods in between, while  for small populations--but above the threshold for the stability and relevance of the endemic state--one observes frequent outbreaks, of small intensity (on the scale of the population size), while the number of infected individuals does not fluctuate too much.
When overpopulation should be taken into account, the explicit solution of ref.~\cite{frame1974explicit} is no longer valid, since a conserved quantity no  longer exists.  Numerical solutions may still, of course, be obtained and do not indicate anything especially interesting happening: the solutions are, of course, no longer periodic and the amplitudes decrease with time. 

\begin{figure}[thp]
\begin{center}
\includegraphics[scale=0.5]{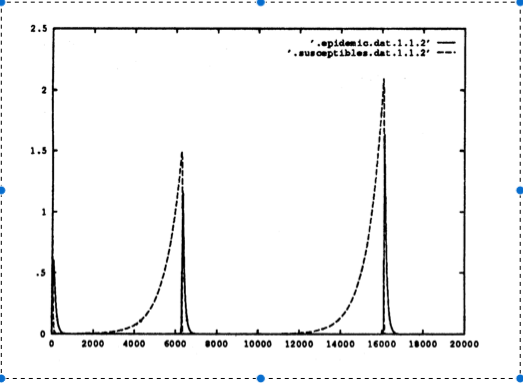}
\end{center}
\caption[]{\sl Time evolution of susceptibles $x(t)$ (dashed line) and infectives $y(t)$ (solid line) for $1/\mu=75$ days, $1/\gamma=10$ days, $\beta_\infty=1$, $\kappa=1000$ and a population of $N=1000$; $T=120$ days and $\epsilon=0.8$.  }
\label{Fig3}
\end{figure}

\begin{figure}[thp]
\begin{center}
\includegraphics[scale=0.5]{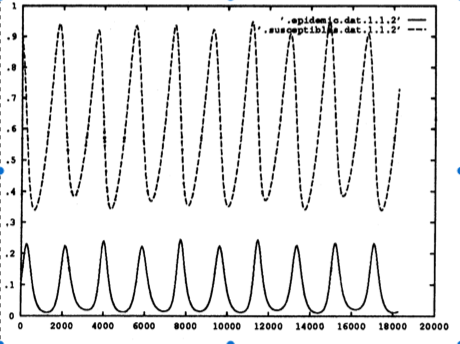}
\end{center}
\caption[]{\sl  Time evolution of susceptibles $x(t)$ (dashed line) and infectives $y(t)$ (solid line) for $1/\mu=75$ days, $1/\gamma=10$ days, $\beta_\infty=1, \kappa=1000$ and a population of $N=200$; $T=120$ days and $\epsilon=0.4$.}
\label{Fig4}
\end{figure}

Finally, an issue that has attracted some attention in the literature is that of the {\em seasonal variation} of the infection rate. One assumes that the infection rate $\beta$ is a periodic function of time, with a period of several months. The typical form is 
\begin{equation}
\label{seasonalvarinf}
\beta_s=\beta\left(1+\epsilon\sin\frac{2\pi}{T}t\right)
\end{equation}
with $\beta$ given by eq.~(\ref{finitebetaN}) and $\epsilon$ a constant between zero and 1 (so that the infection rate remain always positive!). In figs.~\ref{Fig3},~\ref{Fig4} 
we show some examples of what can happen to the cases of figs.~\ref{Fig1},~\ref{Fig2}.  With a system of two ordinary differential equations, the most exciting thing that may occur is a transition from periodic to quasiperiodic behavior--as my be seen--but no chaos, for which at least three equations are needed; for an example see ref.~\cite{metz1990childhood}.

\subsection{The reduced S--I--R model}\label{reducedSIR}
A very similar model was studied in ref.~\cite{schenzle1987critical}.  It is a simplification of the full model discussed in the opening paragraphs (cf. eq.~(\ref{SIRmodel})) and its equations may be cast in the following form
\begin{equation}
\label{reducedSIRmodel}
\begin{array}{lcc}
\displaystyle
dx/dt & = & \displaystyle -\beta(N)xy+\mu(1-x)\\
\displaystyle
dy/dt & = & \displaystyle \beta(N)xy-\gamma y
\end{array}
\end{equation}
This entails a change in the physical meaning of the parameter $\mu$: whereas in the Lotka--Volterra model it contributed a term $+\mu x$, and was interpreted as a birth (or renewal) rate, it now contributes a term $-\mu x$ and should be interpreted as a {\em removal} (or death) rate of the susceptibles. Furthermore, there is a constant source term proportional to $\mu$, to `counterbalance' the two loss terms. The analysis of the original system may easily be carried over. The fundamental difference is that this system does not exhibit oscillatory behavior~\cite{sel1968self}.

\begin{figure}[thp]
\begin{center}
\includegraphics[scale=0.5]{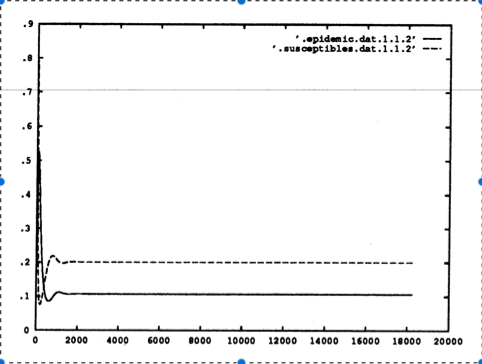}
\end{center}
\caption[]{\sl Time evolution of susceptibles $x(t)$ (dashed line) and infectives $y(t)$ (solid line) for $1/\mu=75$ days, $1/\gamma=10$ days, $\beta_\infty=1, \kappa= 1000$ and a population of $N=1000$. Time is in days.}
\label{Fig5}
\end{figure}

\begin{figure}[thp]
\begin{center}
\includegraphics[scale=0.5]{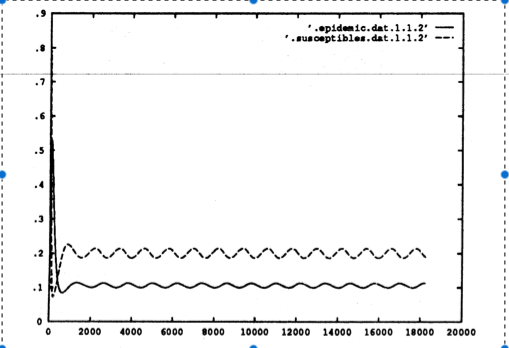}
\end{center}
\caption[]{\sl Time evolution of susceptibles $x(t)$ (dashed line) and infectives $y(t)$ (solid line) for $1/\mu=75$ days, $1/\gamma=10$ days, $\beta_\infty=1, \kappa=1000, \epsilon=0.05$ and a population of $N=1000$. Time is in days. }
\label{Fig6}
\end{figure}

There are two stationary configurations
\begin{itemize}
\item {\em Immune State}. $(x^\ast,y^\ast)=(1,0)$. The eigenvalues of the stability matrix are $\xi_1=\beta-\gamma$ and $\xi_2=-\mu$; it will be stable to small fluctuations if $\beta < \gamma$, or
\begin{equation}
\label{NcritimmunredSIR}
N < N''_\mathrm{cr}=\kappa\frac{\gamma}{\beta_\infty-\gamma}
\end{equation}
\item {\em Endemic state}: $(x^\ast,y^\ast)=(\gamma/\beta,(\mu/\gamma)(1-\frac{\gamma}{\beta})$. For this state to be biologically relevant one must have 
\begin{equation}
\label{NcritendemicreducedSIR}
\kappa\frac{\gamma}{\beta_\infty-\gamma} < N < \kappa\frac{1}{\frac{\beta_\infty(\mu-\gamma)}{\mu\gamma}-1}
\end{equation}
when the denominator is positive; otherwise the point is always relevant. For it to be stable under small fluctuations one must violate constraint~(\ref{NcritimmunredSIR}).

For the values used in the previous model we find that 
\begin{equation}
\label{Nbounds}
N_\mathrm{cr}=128 < N=1000
\end{equation}
and that the denominator is negative, i.e. the endemic state of this model is stable. In fig.~\ref{Fig5} we display the time evolution of the densities $x(t)$ and $y(t)$ for typical values of the parameters.
\end{itemize}
Seasonal variations of the infection rate may also be included in this model. The system, by itself, does not possess an oscillatory state; when driven, however, by an oscillatory infection rate, it {\em does} go into such a state, if the coefficient $\epsilon$ is large enough. In fig.~\ref{Fig6} we show an example.
We now turn to the probabilistic formulation of the spread of an epidemic.

\section{The Stochastic Approach}\label{stochapp}
A consistent treatment of the fluctuations due to the finite size of the population requires a stochastic approach. The aim of the exercise is to write evolution equations for the probability $p_{x,y}(t)$ of finding a fraction $x$ of susceptibles and a fraction $y$ of infected individuals at time $t$. These equations have been considered since the very beginning of the theory for infectious diseases (cf.~\cite{bailey1975mathematical,van1984stochastic})--but their analytical complexity led to the study of approximation schemes that were more transparent. 

This effort resulted in the `stochastic threshold theorems'~\cite{bailey1975mathematical}: these state that an epidemic may become endemic if the susceptible population exceeds a critical size if the population is `large enough', with probability 1. It was later possible to relax considerably the somewhat vague constraint on the population size~\cite{ reinert1989}. We shall present here a computer--aided procedure that allows one to attack the original problem directly and obtain, in some cases, exact expressions for the $p_{x,y}(t)$. This was, unfortunately, possible only in two limiting cases--the general stochastic epidemic~\cite{bailey1975mathematical} (where infectives are removed and susceptibles are not renewed) and in the case where infectives are not removed but susceptibles are renewed.

\subsection{No renewal of the susceptibles}\label{norenew}
To set the stage we shall first concentrate on the ``general stochastic epidemic'', neglecting renewal and overpopulation (which is not a problem, since the population always decreases).

The evolution equation is derived in the standard way~\cite{bailey1975mathematical}--the assumption here is that within the time interval $(t,t+\Delta t)$ only one infection or removal may take place and it will be convenient to work with population sizes $X=Nx, Y=Ny$ rather than densities
$$
\begin{array}{c}
\displaystyle
p_{X,Y}(t+\Delta t)=\\
\displaystyle
\beta\Delta t(X+1)(Y-1)p_{X+1,Y_1}(t)+\gamma\Delta t(Y+1)p_{X,Y+1}(t)+(1-\beta\Delta t XY-\gamma\Delta t Y)p_{X,Y}(t);
\end{array}
$$
 taking the limit $\Delta t\to 0$ one obtains a (potentially infinite) system of linear ordinary differential equations 
 \begin{equation}
 \label{masterode}
 \begin{array}{c}
 \displaystyle
 dp_{X,Y}/dt = \\
 \displaystyle
 \beta(X+1)(Y-1)p_{X+1,Y-1}(t)+\gamma(Y+1)p_{X,Y+1}(t)-(\beta XY+\gamma Y)p_{XY}(t)
 \end{array}
 \end{equation}
 Let us assume, now, that the maximum number of infected individuals in the population cannot exceed some upper limit $a$ (a reasonable assumption); it is also clear that the number of susceptible individuals cannot increase in time. The simplest situation is when the composition of the population at time $t=0$ is $(n,a), n+a=N$, i.e. $p_{X,Y}(0)=\delta_{X,n}\delta_{Y,a}$. This means that the equation~(\ref{masterode}) for $X=n, Y=a$ yields immediately
\begin{equation}
\label{pna}
p_{n,a}(t)=e^{-a(n+\gamma/\beta)\beta t};
\end{equation}  
while, in principle, this leads to a unique solution for equations~(\ref{masterode}), in practice the procedure is exceedingly cumbersome both for analytical understanding as for practical calculations~\cite{bailey1975mathematical}. It is possible, however, to transform equations~(\ref{masterode}) to an integral recursion relation that is quite transparent
and computationally easy to use; furthermore, it leads to exact expressions for a subset of the quantities $p_{X,Y}(t)$. 

First note that the solution for $p_{X,Y}(t)$ is of the form
\begin{equation}
\label{solAnsatz}
p_{X,Y}(t)=C(X,Y)e^{-(XY+(\gamma/\beta)Y)\beta t};
\end{equation} 
it is useful to rescale time so that $\beta t=\tau$ and set $\rho=\gamma/\beta$. Substituting in eq.~(\ref{masterode}) and using the initial condition, one ends up with 
\begin{equation}
\label{integralrecursion}
\begin{array}{c}
\displaystyle p_{X,Y}(\tau)=e^{-Y(X+\rho)\tau}\times\\
\displaystyle
\int_0^\tau\,dx\,\left[(X+1)(Y-1)p_{X+1,Y-1}(x)+\rho(Y+1)p_{X,Y+1}(x)\right]\,e^{Y(X+\rho)x}
\end{array}
\end{equation}
This expression already provides some information: aside from $p_{n,a}(\tau)$, all other $p_{X,Y}(t)$ are zero at the origin (that is the initial condition), go exponentially at infinity and have a maximum a some intermediate time, $\tau^\ast_{X,Y}$ (for fixed $n$ and $a$ and $\rho$). 

It is possible to write down an exact expression for the probabilities $p_{n,a-k}(\tau)$, for all $k=0,\ldots,a$, using the recursion eq.~(\ref{integralrecursion})
\begin{equation}
\label{solintegralrecursion0}
p_{n,a-k}(\tau)=\left(\frac{\rho}{n+\rho}\right)^k\left(\begin{array}{c} a\\ k\end{array}\right)e^{-(a-k)(n+\rho)\tau}\left(1-e^{-(n+\rho)\tau}\right)^k
\end{equation}
This is very transparent: it is nothing but a binomial distribution with weight $\varpi=e^{-(n+\rho)\tau}$ (up to a non--trivial combinatoric factor), bearing witness to a relation with the random walk. 

It is straightforward to insert equ.~(\ref{solintegralrecursion0}) in equ.~(\ref{integralrecursion}) and successively compute the rest. The simplest way is to evaluate numerically the integral for all times up to $\tau$--the functions to be integrated over are so smooth that the error committed is not very large as a comparison with the direct evaluation of the first few terms shows. 

The limitation of the previous initial condition is, of course, that an epidemic will never break out, since the infectives start at their peak size. Let us, therefore, consider the general initial condition, 
\begin{equation}
\label{generalinitcond}
p_{X,Y}(0)=\delta_{X,N}\delta_{Y,A}
\end{equation}
with $A < a$. This time $p_{n,a}(\tau)\equiv 0$ as are all $p_{n,a-k}(\tau)$ and, working through the hierarchy, all down to $p_{N,A}(\tau)=\exp(-A(N+\rho)\tau)$. It is once again possible to show by induction that 
\begin{equation}
\label{pNAk}
p_{N,A-k}(\tau)=\left(\frac{\rho}{N+\rho}\right)^k\left(\begin{array}{c} A\\ k\end{array}\right)e^{-(A-k)(N+\rho)\tau}\left(1-e^{-(N+\rho)\tau}\right)^k
\end{equation}
and the $p_{N-l,A-k}(\tau)$ may be obtained through the recursion relation eq.~(\ref{integralrecursion}).

It is now possible to  study the outbreak of an epidemic and test another facet of the threshold theorems--an infection will break out only when the number of susceptibles is sufficiently high. It is, in fact, more useful to turn this around and say that an infection will break out when its parameter $\rho$ is sufficiently low--which indicates what kind of populations are prone to be vulnerable to what kind of diseases. 

In the following subsection we compare the predictions of the deterministic and stochastic formulations with regard to this threshold behavior.

\subsection{Threshold behavior}\label{threshbeh}
Let us recall, first of all, what happens in the corresponding deterministic system. Since the susceptibles are not renewed, the equations are
\begin{equation}
\label{threshdet0}
\begin{array}{lcc}
\displaystyle dx/dt & = & -\beta xy\\
\displaystyle dy/dt & = & y(\beta x-\gamma)
\end{array}
\end{equation} 
It is again convenient to rescale time through $\tau=\beta t$ and set $\rho=\gamma/\beta$. It is possible to reduce eqs.~(\ref{threshdet0}) to
\begin{equation}
\label{threshdet1}
\begin{array}{ccccc}
\displaystyle y+x-\rho\ln\,x & = & \displaystyle y_0+x_0-\rho\ln\,x_0 & = & C\\
\displaystyle dx/d\tau         & = & \displaystyle -x(C-x+\rho\ln\,x) & & \\
\end{array}
\end{equation}
It is easy to see that the density of infected individuals $y$ has an extremum for $x=\rho$ and to check that this extremum is a mximum ($d^2y/d\tau^2=-\rho y^2 <0$). The number of infected individuals at this peak is
\begin{equation}
\label{peakinfectives}
y_p = C-\rho+\rho\ln\,\rho
\end{equation}
Since the number of susceptibles decreases with time, this peak will be accessible only if $x_0 > \rho$, expressing the content of the deterministic threshold theorem. Furthermore, this peak will be an outbreak if $y_p>y_0$--this last condition may also be expressed as a threshold on $x_0$. The dependence on the population size is carried by $\rho$ through the infection rate $\beta$. 

It is also possible to evaluate to calculate the total size of the epidemic, i.e. how many susceptible individuals become infected in the end: after a very long time, the number of  
number of infected individuals goes to zero, since they are removed at a rate $\gamma$ and the susceptibles do not reproduce. The density of susceptibles at that stage is given by the solution of
\begin{equation}
\label{xinfty}
C = x_\infty-\rho\ln\,x_\infty
\end{equation}
 and the total size of the epidemic is 
 \begin{equation}
 \label{sizeepid}
 S = x_0-x_\infty
 \end{equation}
Since these are `physically relevant' quantities it is instructive to see what may be deduced from the corresponding stochastic model. The analog of the peak size of the epidemic, $Y_p=y_pN$ is
\begin{equation}
\label{stochpeaksize}
\begin{array}{c}
\displaystyle Y_p=\left\langle Y\right\rangle(\tau');\\
\displaystyle \left.\frac{\partial\left\langle Y\right\rangle(\tau)}{\partial\tau}\right|_{\tau=\tau'}=0
\end{array}
\end{equation} 
where $\langle\ldots\rangle$ denotes the average with respect to the probabilities $p_{X,Y}(\tau)$. In a similar fashion it is possible to compute the average size of the epidemic. The problem here is that this last quantity is much more susceptible to numerical error since it involves the computation of exponentially small quantities 
($e^{-\mathrm{const}\, t}$ for large values of $t$). What one finds is that the sharp threshold behavior is smeared out for finite populations and that the critical region around the expected value $\rho_c=n$ is quite wide.

\subsection{Contribution of a flux of susceptibles}\label{fluxsusc}
Let us now consider the case, where individuals enter the susceptible population at a rate $\mu$; eq.~(\ref{masterode}) is then modified by the presence of an extra term
\begin{equation}
\label{masterodefluxsusc}
\begin{array}{c}
\displaystyle
dp_{X,Y}/dt=\\
\displaystyle
\beta(X+1)(Y-1)p_{X+1,Y-1}(t)+\mu(X-1)p_{X-1,Y}(t)+\gamma (Y+1)p_{X,Y+1}(t)-(\beta XY+\gamma Y+\mu X)p_{X,Y}(t)
\end{array}
\end{equation} 
--this modification ignores overpopulation effects, though including them, through a term $\mu\nu X^2$ is immediate. It complicates the analysis in no mean fashion, since the system is no longer ``naturally'' closed; the population may grow without bound, since it is continuously renewed; but the fundamental difficulty stems from the competition between the renewal of the susceptibles and the removal of the infectives. 

If, however, we consider the case where the infected individuals are {\em not} removed from the population (i.e. $\gamma=0$) then it is possible to obtain an exact solution. 
 This is due to the fact that the equation for $p_{1,1}$ may be solved immediately.
 
 In fact, id one chooses as initial condition $p_{X,Y}(0)=\delta_{X,1}\delta_{Y,1}$ one may obtain the exact solution of eq.~(\ref{masterodefluxsusc}) as follows (setting $\tau=\beta t$ and $\sigma=\mu/\beta$).
 \begin{itemize}
 \item The expression for the $p_{k,1}$, $k=1,\ldots$ is 
 \begin{equation}
 \label{pk1}
 p_{k,1}(\tau)=\left(\frac{\sigma}{1+\sigma}\right)^{k-1}e^{-k(1+\sigma)\tau}(e^\tau-1)^{k-1}
 \end{equation}
 \item The remaining $p_{X,Y}$'s may be obtained as follows: for a given $X$ one computes the $p_{X,Y}$ from the $p_{X,Y-1}$ through the integral recursion relation
 \begin{equation}
 \label{newintegralrecurs}
 p_{X,Y}(\tau)=e^{-X(Y+\sigma)\tau}\int_0^\tau\,dx\,\left[ (X+1)(Y-1)p_{X+1,Y-1}(x)+\sigma(X-1)p_{X-1,Y}(x)\right]\,e^{X(Y+\sigma)x}
 \end{equation}
 For the general initial condition $p_{X,Y}(0)=\delta_{X,N}\delta_{Y,A}$ the solution is 
 \begin{equation}
 \label{solex}
 p_{N+k,A}(\tau)=\left(\frac{\sigma}{A+\sigma}\right)^k\left(\begin{array}{c} N+k-1\\ k\end{array}\right)\left( e^{A+\sigma)\tau}-1\right)^k e^{-(N+k)(A+\sigma)\tau}
 \end{equation}
 with the rest obtained from recursion~(\ref{newintegralrecurs}), while $p_{X<N,Y<A}(\tau)\equiv 0$. 
 
 Eq.~(\ref{masterodefluxsusc}) is a stochastic version of the first model studied previously; in a similar vein one may write down the stochastic version of the second model, where susceptibles both entered and left the population
 \begin{equation}
 \label{generalcase}
 \begin{array}{c}
 \displaystyle
 dp_{X,Y}/dt = \\
 \displaystyle
 \beta(X+1)(Y-1)p_{X+1,Y-1}(t)+\mu(X+1)p_{X+1,Y}(t)+\mu (X-1)p_{X-1,Y}(t)+\gamma(Y+1)p_{X,Y+1}(t)-\\
 \displaystyle
 (\beta XY+\gamma Y+\mu (X+1))p_{X,Y}(t)
 \end{array}
 \end{equation}
 \end{itemize}
 \section{Conclusions}\label{concl}
 The purpose of the present study was to analyze in detail a particular assumption of how the infection rate depends on population size. Within the deterministic approximation
  we found intervals for the relevant parameter values, where the disease becomes endemic. In particular, we showed that the parameter $\kappa$, that controlled the finite--size effects, controls in a decisive manner whether the theoretical constraints may be of any practical relevance. Furthermore, by direct numerical monitoring of the evolution equations we could check how the values of the other parameters, such as the renewal rate of the susceptibles  and the removal rate of the infectives affected other features of the model, especially oscillatory behavior. In this regard we examined two special models; one was a typical Lotka--Volterra model, the other a model typcially used for studying epidemic spread, belonging to the so--called S-I-R class~\cite{metz1990childhood}. In the first case we found that the characteristic period of the oscillations is always much larger than the characteristic times (renewal rate, duration of the disease). This places constraints on the range of applicability of the model to certain cases claimed previously: the renewal rate of the susceptibles, in particular, must {\em not} be identified with a life expectancy; or, if it should, then the model is simply not relevant for human populations. 
   It is especially relevant, though, for cases of highly mobile populations; and many human communities do satisfy this condition. An interesting prediction is that diseases of fairly short duration, in populations with high renewal rate, lead to oscillations with a much longer period; also that large, infrequent, outbreaks occur in populations large, compared to the parameter $\kappa$, whereas small populations show small, infrequent ones.
   
   If one, 	additionally, assumes a seasonal variation of the infection rate, 	a transition to quasiperiodic behavior is observed. 
   
   In the second model we established new bounds on the population size for the disease to become endemic. We recovered the result~\cite{metz1990childhood} that seasonal variation of the infection rate leads, beyond a certain threshold, to oscillatory behavior in this case also. The remarks on the interpretation of the renewal rate (here a removal rate) of the susceptibles set forward for the first model apply here as well; both models have as natural field of application highly mobile communities, where the epidemic breaks out on time scales much longer than the typical residence times of an individual in the system. This indicates that eradication policies have a narrow window of applicability. 
   
   Finally we presented a stochastic formulation.  This is useful since the assumptions, on which the deterministic approximation rests, are not always valid, in particular for small populations. For the classical problem of the general stochastic epidemic, we obtained a computer--aided exact solution, that allows us to compute the transition probabilities exactly. This procedure is particularly well--suited for small populations--and time scales of the same order of magnitude as the characteristic times of the population; for large ($N>40$) and very long times rounding errors are exceedingly important and the extraction of reliable results quite delicate. Our explicit expressions may allow checks and direct computations through symbolic manipulation progams. 
   
   Regarding possible extensions of the present formalism: while we have taken fluctuations into account, within the stochastic models, the present approach retains a `mean--field' character, since any infected individual may infect any susceptible one; spatial fluctuations are ignored and work in this direction remains to be done. 
   
   It might also be possible to study the `intermediate region', between the two limiting cases that have been solved here, to wit, when both susceptibles are renewed and infectives are removed. 
   
   Regarding the deterministic approximation, the full model remains a challenge; for it to be meaningful, however, one needs explicit expressions for the functions that control the finite--size effects. 
   
   {\bf Acknowledgements}: We would like to thank Prof. Dr. K. Dietz for very helpful criticism. S.N. would also like to acknowledge fruitful discussions with H. J. Herrmann, G. Nicolis, V. Privman and N. \v{S}vraki\'c.    
\bibliographystyle{utphys}
\bibliography{BiBoS}

\end{document}